\documentclass[aip, jcp,preprint,showpacs,superscriptaddress,groupedaddress]{revtex4-1}

\usepackage{amsmath}
\usepackage{amssymb}
\usepackage{braket}
\usepackage{bm}

\usepackage{graphicx}
\usepackage{subcaption}
\usepackage{mwe}
\usepackage{color}

\newcommand*{\thead}[1]{\multicolumn{1}{c}{#1}}

\begin{document}

\title{Richardson-Gaudin mean-field for strong correlation in quantum chemistry}
\author{Paul A. Johnson}
 \email{paul.johnson@chm.ulaval.ca}
 \author{Charles-\'Emile Fecteau}
 \author{Fr\'ed\'eric Berthiaume}
 \author{Samuel Cloutier}
 \author{Laurie Carrier}
 \author{Marianne Gratton}
 \affiliation{D\'{e}partement de chimie, Universit\'{e} Laval, Qu\'{e}bec, Qu\'{e}bec, G1V 0A6, Canada}

\author{Patrick Bultinck}
 \affiliation{Department of Chemistry, Ghent University, Krijgslaan 281 (S3), 9000 Gent, Belgium}

\author{Stijn De Baerdemacker}
 \affiliation{Department of Chemistry, University of New Brunswick, 30 Dineen Drive, Fredericton, New Brunswick, E3B 5A3, Canada}
\author{Dimitri Van Neck}
 \affiliation{Center for Molecular Modeling, Ghent University, Technologiepark 903, 9052 Zwijnaarde, Belgium}

\author{Peter Limacher}
\author{Paul W. Ayers}
 \affiliation{Department of Chemistry and Chemical Biology, McMaster University, Hamilton, Ontario, L8S 4M1, Canada}

\date{\today}

\begin{abstract}
Ground state eigenvectors of the reduced Bardeen-Cooper-Schrieffer Hamiltonian are employed as a wavefunction ansatz to model strong electron correlation in quantum chemistry. This wavefunction is a product of weakly-interacting pairs of electrons. While other geminal wavefunctions may only be employed in a projected Schr\"{o}dinger equation, the present approach may be solved variationally with polynomial cost. The resulting wavefunctions are used to compute expectation values of Coulomb Hamiltionans and we present results for atoms and dissociation curves which are in agreement with doubly-occupied configuration interaction (DOCI) data. The present approach will serve as the starting point for a many-body theory of pairs, much as Hartree-Fock is the starting point for weakly-correlated electrons.
\end{abstract}

\maketitle

\section{Introduction}

The majority of electronic structure methods are built upon the orbital picture. In the simplest models, electrons are understood to behave essentially independently, interacting only with the average field produced by the other electrons. This picture is acceptable when it is possible to assign orbitals unambiguously as occupied or unoccupied, i.e., when the energy gap between the occupied and unoccupied orbitals is large compared with the kinetic energy of the valence electrons. The simplest wavefunction ansatz, i.e. a single Slater determinant with a set of $N$ (number of electrons) occupied orthonormal spin orbitals optimised to produce the lowest energy in Hartree-Fock theory already produces a good first approximation. Further expansion in terms of Slater determinants is obtained by including singly-, doubly-, ... excited Slater determinants with respect to the Hartree-Fock Slater determinant leading to a better approximation of the electronic wavefunction that is most often dominated by the Hartree-Fock determinant \cite{helgaker_book}. Systems of this type are \emph{weakly-correlated} and are generally well described by density functional theory (DFT), and coupled-cluster theory\cite{bartlett:2007} with singles, doubles, and (perturbative) triples. 

However, when it is difficult to label orbitals as occupied or unoccupied, this picture breaks down. The number of important Slater determinants grows exponentially with the system size, so a single Slater determinant is not a qualitative representation of the electronic wavefunction. Such systems are \emph{strongly-correlated}. State of the art methods include the density matrix renormalization group (DMRG)\cite{RN1776,RN1779,RN1780,RN1781,RN1782,RN1783,RN1785,RN1786,RN1787,RN1788,RN1789,RN1790,RN1791,RN1792} and Slater determinant Monte-Carlo \cite{RN1793,RN1794,RN1795,RN1796,RN1797,RN1798,RN1799,RN1800,RN1801,RN1802,RN1804,RN1805,RN1806,RN1807}. The high computational cost of these methods has motivated the pursuit of approximate methods for treating strongly-correlated systems generally with mean-field cost. This contribution is a step in that direction: we employ the eigenvectors for a schematic system as a variational ansatz. The key idea is to work in a framework in which strong-correlation is described by the mean-field. In this new basis, the electronic wavefunction will have a short expansion dominated by a single contribution, though it will not necessarily be a Slater determinant.

Thus, we have been studying wavefunctions built as products of geminals, i.e. pairs of electrons. 
Geminal wavefunctions have been proposed since the founding days of quantum chemistry, as they tie in with the intuitive chemical picture of Lewis bonds as pairs of electrons \cite{hurley:1953}. Unfortunately, the most general geminal wavefunctions come with a computational cost that scales exponentially with the system size, hence they were soon abandoned for other methods. The intrinsic reason for the pernicious cost can be inferred from the antisymmetrized product of interacting geminals (APIG), which is a general product of closed-shell singlet geminals \cite{silver:1969}. 

The expansion coefficients in the basis of Slater determinants are \emph{permanents} of the geminal coefficients, and are combinatorially difficult to evaluate in general \cite{minc:1978}. There are several approaches to simplifying the problem to one that is tractable, each of which amounts to making particular approximations. The first is to make all the geminals identical resulting in the antisymmetrized geminal power (AGP)\cite{coleman:1997}, or equivalently a number-projection of the Bardeen-Cooper-Schrieffer (BCS) ansatz\cite{bardeen:1957,ring:1980}. AGP is well studied and is easy to employ, but suffers the major drawback of not being size-consistent. The second is to partition the orbitals in such a way that in each geminal, each orbital has one partner as in generalized valence bond-perfect pairing (GVB-PP)\cite{goddard:1967}, a set of unique partners as in the antisymmetrized product of strongly-orthogonal geminals (APSG)\cite{surjan:1999}, or one major occupied contribution as in the antisymmetrized product of 1-reference orbital geminals (AP1roG\cite{peter}). In a series of papers, we have proposed and investigated AP1roG as a computationally facile wavefunction to describe strong correlation due to bond-breaking\cite{johnson:2013,peter,boguslawski:2014a,boguslawski:2014b, boguslawski:2014c, tecmer:2014}. It was found that AP1roG systematically reproduces ground-state energies of doubly-occupied configuration interaction (DOCI) calculations for molecular systems\cite{peter}, even for large system sizes\cite{shepherd:2014}. The key ingredient in the AP1roG formalism is that the Schr\"{o}dinger equation is solved projectively with respect to a set of selected reference states, very much in the spirit of coupled-cluster theory (CC). In particular, it is equivalent to pair-coupled-cluster doubles (pCCD)\cite{pCCD,henderson:2014a,henderson:2014b,bulik:2015}. As a result, the permanents one needs to compute all become very easy to evaluate. Indeed, the computational bottleneck in the AP1roG calculations is the orbital optimization (OO), rather than the computation of the geminal coefficients in the AP1roG wavefunction. The energies of geminal theories are strongly dependant on the orbital pairing scheme used\cite{boguslawski:2014c,limacher:2014a}. It was found that Hartree-Fock orbitals are typically well-suited for pairing correlations around equilibrium geometries, whereas pairing occurs preferentially in localized orbitals in the bond-dissociation regime\cite{limacher:2014a}. Geminal theories perform well in the latter regime as strong correlations tend to dominate weak ones. Efforts to incorporate weak correlation in these geminal wavefunctions rely on multi-reference pertubation theory (MRPT)\cite{kobayashi:2010,limacher:2014b}, the random-phase approximation (RPA)\cite{pastorczak:2015}, or generalize the algebraic structure of spin-singlet geminal theory in order to include spin-triplet excitations as well\cite{johnson:2017}. The main hurdle on the way to an all-inclusive geminal theory is the ability to include missing correlations systematically, like CC theory or truncated configuration-interaction (CI) approaches\cite{helgaker_book}. This is one of the main motivations for the approach herein presented. It has recently been observed that the \emph{seniority} scheme provides a new means to organize the Hilbert space for configuration interaction approaches in a hierarchical way. The seniority quantum number counts the number of electrons that are \emph{not} paired\cite{bytautas:2011}. In this framework, the DOCI method corresponds to the seniority-zero rung on the ladder. DOCI is size-extensive, in the correct basis, and captures the majority of strong-correlation, at the cost of combinatorial scaling typical for full CI methods albeit now in pair space.

With this in mind, we follow a third approach in this contribution, in which we propose a structured geminal wavefunction such that the required permanents may be easily evaluated. Specifically, we employ the ground-state eigenvectors of the reduced BCS Hamiltonian, or Richardson Hamiltonian\cite{RN1587} as a variational wavefunction ansatz. It is well-established that the Richardson, or Richardson-Gaudin (RG), model is a quantum integrable system for which the eigenvectors can be obtained using Bethe-Ansatz techniques\cite{dukelsky:2004,ortiz:2005}. From a quantum chemistry point of view, it is highly remarkable that these eigenvectors have the structure of a geminal wavefunction, completely characterized by means of the single-electron model parameters, the pairing strength, and a set of so-called rapidities. There are as many rapidities as there are electron pairs which means that the eigenvectors can be determined by solving a set of equations for the rapidities with a computational scaling that is linear with the system size, rather than the typical combinatorial scaling. Moreover, the norms, scalar products, and 1- \& 2-body reduced density matrices (1-RDM \& 2-RDM) can all be computed with a polynomial cost. This opens an avenue for a variational geminal theory in quantum chemistry. We have already reported first results for LiH, Li$_2$ and HF dissociation curves in previous work\cite{tecmer:2014}, so we will focus on the mathematical details in the present paper.

It should be emphasized that in our approach the object being optimized is the model Hamiltonian (in this case RG), not simply an ansatz for the wavefunction. As a result, we obtain a complete set of eigenvectors with which to construct perturbative corrections, Green's functions etc., all of which are physically well-founded and interpretable. 

Similar ideas using the framework of exactly-solvable models as a many-body expansion technique are being explored outside the field of quantum chemistry. In nuclear structure physics, eigenvectors of RG models are being used as a starting point for a CI approach\cite{debaerdemacker:2017}. In condensed matter physics, a variational RG approach is used to treat integrability-breaking interactions in central-spin problems\cite{claeys:2017a}, and a CI framework has been developed for non-integrable spin chains in the truncated spectrum approximation\cite{james:2018}. Recently, we have developed the analogue of Hartree-Fock as a Bethe ansatz to serve as a bridge to the present contribution\cite{laurie}.

In section \ref{sec:ansatz} we outline the basics of RG models, introduce the eigenvectors and develop the variational principle to be employed. In section \ref{sec:numbers} we present numerical results for 4-, 6-, 8- and 10-electron atomic systems as well as dissociation curves for H$_2$, H$_4$, H$_6$, H$_8$ and N$_2$. We formulate our conclusions in section \ref{sec:conclusions}.

\section{Variational Ansatz} \label{sec:ansatz}

\subsection{Eigenvectors of the Reduced BCS Hamiltonian}

We employ a pseudospin representation of su(2) for a set of spatial orbitals $\{i\}$, each of which can contain a single pair of opposite spin electrons. For each spatial orbital there are three operators:
\begin{align}
S^+_i = a^{\dagger}_{i\uparrow} a^{\dagger}_{i\downarrow}, \quad S^-_i = a_{i\downarrow}a_{i\uparrow}, \quad S^z_i = \frac{1}{2}\left( a^{\dagger}_{i\uparrow}a_{i\uparrow} + a^{\dagger}_{i\downarrow}a_{i\downarrow} -1 \right),
\label{eq:pseudospin}
\end{align}
where $a^{\dagger}_{i\uparrow} \; (a_{i\downarrow})$ creates (removes) an up- (down-)spin electron in spatial orbital $i$, etc. $S^+_i$ adds a pair of electrons to spatial orbital $i$ and $S^-_i$ removes a pair. Each spatial orbital can only hold one pair. Acting on a doubly-occupied spatial orbital, $S^z_i$ gives $\frac{1}{2}$, while acting on an empty spatial orbital $S^z_i$ gives $-\frac{1}{2}$. Thus $S^z_i$ ``measures'' whether spatial orbital $i$ is full or empty. For singly occupied orbitals, its action is zero. For doubly-degenerate spin orbitals, the seniority quantum number can be obtained as the expectation value of the operator \cite{bytautas:2011,alcoba:2014}
\begin{align}
\Omega_i = a^{\dagger}_{i\uparrow}a_{i\uparrow} + a^{\dagger}_{i\downarrow}a_{i\downarrow} - 2a^{\dagger}_{i\uparrow}a_{i\uparrow}a^{\dagger}_{i\downarrow}a_{i\downarrow}.
\end{align} 
We work only in the seniority zero sector $\braket{\Omega_i}=0,\;\forall i$ in this paper, although it is perfectly possible to extend the formalism to other sectors. This extension is one of the strengths of the present approach.

Using the fermionic anticommutation relations, it is easily verified that the operators, $\eqref{eq:pseudospin}$, commute for distinct spatial orbitals, so that the structure constants of their Lie algebra may be summarized
\begin{align}
\left[S^z_i, S^{\pm}_j\right] = \pm \delta_{ij}S^{\pm}_i, \quad
\left[S^+_i, S^-_j\right]     = 2 \delta_{ij}S^z_i.
\end{align}
With $\hat{n}_i = 2S^z_i +1$, which counts the electrons in spatial orbital $i$, the reduced BCS Hamiltonian\cite{RN1587,RN1171,RN810} for a system with $K$ spatial orbitals is
\begin{align}
\hat{H}_{BCS} = \frac{1}{2}\sum^{K}_{i} \varepsilon_i \hat{n}_i - g \sum^K_{ij} S^+_i S^-_j,
\end{align}
where the parameters defining the system are the single particle spectrum $\{\varepsilon_i\}$, and the pairing strength $g$. In this convention, a positive $g$ represents an attractive pairing interaction. The eigenvectors are products of electron pairs distributed over the entire space of orbitals, each pair being characterized by a complex number $u$ (which we call a rapidity). Such an electron pair is denoted
\begin{align}
\mathbb{S}^+(u) = \sum^K_{i} \frac{S^+_i}{u - \varepsilon_i},
\end{align}
and for a system with $M$ pairs of electrons, the states
\begin{align}
\ket{\{ u \}} &= \prod^{M}_{a} \mathbb{S}^+(u_a) \ket{\theta}
\label{eq:ABA}
\end{align}
are eigenvectors of the reduced BCS Hamiltonian provided the rapidities satisfy a set of coupled non-linear equations, called Richardson's equations
\begin{align} \label{eq:RichEquations}
\frac{2}{g} +\sum^K_{i}\frac{1}{u_a -\varepsilon_i} + \sum^M_{b\neq a} \frac{2}{u_b -u_a} =0, \quad \forall\; a=1,M.
\end{align}
The state $\ket{\theta}$ is the vacuum, defined such that 
\begin{align}
S^-_i \ket{\theta} = 0, \quad \forall \;i=1,K
\end{align}
meaning that it is destroyed by all pair removal operators. In this contribution we take $\ket{\theta}$ to be the empty state, but it could easily be taken as a set of non-interacting unpaired electrons (Slater determinant). The reduced BCS Hamiltonian was first solved by Richardson\cite{RN1587,RN1171} and elaborated by Gaudin \cite{RN810}. Thus as a shorthand, we will refer to the state \eqref{eq:ABA} as a Richardson-Gaudin (RG) state.

For a system with $M$ pairs distributed among $K$ spatial orbitals, there are $\binom{K}{M}$ eigenvectors corresponding to the $\binom{K}{M}$ solutions of Richardson's equations. These equations are highly non-linear, with singularities hampering a straightforward numerical characterization of the eigenvectors. In the last decade, many new numerical methods have been developed to properly control and possibly avoid the singularities in the equations. These methods range from clusterization methods \cite{rombouts:2004}, Heine-Stieltjes connections\cite{guan:2012}, probabilistic approaches\cite{pogosov:2012}, pseudo-deformations of the su(2) pairing algebra \eqref{eq:pseudospin}\cite{RN1448} and, most recently, eigenvalue-based methods\cite{faribault:2011,claeys:2015}.In this work we employed eigenvalue-based methods.  

\subsection{Energy Functional}
We will now outline a variational method employing the eigenvectors of the reduced BCS Hamiltonian (as a model system) to approximate solutions of a Coulomb Hamiltonian describing physical electrons. The reduced BCS Hamiltonian is defined by $K+1$ parameters: the single particle energies $\{ \varepsilon\}$ and the pairing strength $g$, which are the variational parameters. Our purpose is to employ the RG state $\ket{\{u\}}$ as a variational ansatz for a Coulomb Hamiltonian,
\begin{align}
\hat{H}_C &= \sum^K_{ij} h_{ij} \sum_{\sigma} a^{\dagger}_{i\sigma}a_{j\sigma} + \frac{1}{2} \sum^K_{ijkl} V_{ijkl} \sum_{\sigma\tau} a^{\dagger}_{i\sigma}a^{\dagger}_{j\tau}a_{l\tau}a_{k\sigma},
\end{align}
with $\sigma$ and $\tau$ spin variables. The one-electron $h_{ij}$ and two-electron integrals $V_{ijkl}$ are calculated in a basis of orthonormal spatial orbitals $\{\phi\}$:
\begin{align}
h_{ij} & = \int d\mathbf{r} \; \phi^*_i(\mathbf{r}) \left( -\frac{1}{2} \nabla^2 -\sum_I \frac{Z_I}{\vert \mathbf{r} -\mathbf{R}_I \vert} \right) \phi_j (\mathbf{r}) \\
V_{ijkl} &= \int d\mathbf{r}_1 d\mathbf{r}_2 \frac{\phi^*_i(\mathbf{r}_1)\phi^*_j(\mathbf{r}_2)\phi_k(\mathbf{r}_1)\phi_l(\mathbf{r}_2)}{\vert \mathbf{r}_1 - \mathbf{r}_2 \vert}
\end{align}
with \textbf{R}$_I$ and $Z_I$ being the positions and charges of the nuclei. 

Thus, with the RG state $\ket{\{u\}}$ as an ansatz, our approximation to the ground state energy is
\begin{align}
E[\{\varepsilon\},g] &= \min_{\{\varepsilon\},g} \frac{\braket{\{u\}|\hat{H}_C|\{u\}}}{\braket{\{u\}|\{u\}}}
\label{eq:energy_functional}
\end{align}
The RG state is the ground state of the reduced BCS Hamiltonian, which is in turn defined by the parameters $\{\varepsilon\}$ and $g$. Thus the energy is to be minimized over these parameters. We do not optimize over rapidities, as they are dictated as the solutions of Richardson's equations for a set of $\{\varepsilon\},g$. It is of paramount importance that we may evaluate \eqref{eq:energy_functional} with a reasonable cost. This is possible thanks to the structure of \eqref{eq:ABA}. The 1-body reduced density matrix (1-RDM) is diagonal and doubly-degenerate, as the $\alpha$ and $\beta$ electrons are treated identically. We adopt the convention
\begin{align}
\gamma_i = \frac{1}{2} \braket{ \{u\} | \hat{n}_i | \{u\} } = \braket{ \{u\} | S^z_i | \{u\} } \label{eq:d1d}
+ \frac{1}{2} \braket{\{u\} | \{u\}}.
\end{align}
Here, $\hat{n}_i$ counts the number of electrons in the spatial orbital $i$, so the elements of the 1-RDM count the number of pairs in each site. They are real numbers between zero and one. The 2-body reduced density matrix (2-RDM) has two non-zero pieces: the \emph{pair correlation function},
\begin{align}
P_{ij} = \braket{\{u\} | a^{\dagger}_{i\uparrow} a^{\dagger}_{i\downarrow} a_{j\downarrow} a_{j\uparrow} | \{u\} } = \braket{ \{u\} | S^+_i S^-_j | \{u\} } \label{eq:d2p}
\end{align}
and the \emph{diagonal correlation function}
\begin{align}
D_{ij} = \frac{1}{4}\braket{ \{u\} | \hat{n}_i\hat{n}_j | \{u\} } = \braket{ \{u\} | S^z_i S^z_j | \{u\} } + \frac{1}{2} \gamma_i + \frac{1}{2} \gamma_j - \frac{1}{4} \braket{\{u\} | \{u\}} . \label{eq:d2d}
\end{align}
The diagonal elements $P_{ii}$ and $D_{ii}$ correspond to the same elements of the 2-RDM, so to avoid double counting we set the elements $D_{ii} = 0$. The state \eqref{eq:ABA} is not normalized, and hence neither are the expressions for the correlation functions \eqref{eq:d1d}, \eqref{eq:d2p}, and \eqref{eq:d2d}

The energy expression becomes:
\begin{align} \label{eq:su2EnergyExpression}
E \braket{ \{u\} | \{u \} }&= 2\sum^K_{i}h_{ii} \gamma_i +\sum^K_{ij} [(2V_{ijij}-V_{ijji})D_{ij} + V_{iijj}P_{ij}]
\end{align}
where the summations are performed over only the spatial orbital index.

The norm and correlation functions of \eqref{eq:ABA} are discussed in refs:\cite{RN1171,RN1643,RN1355,RN1586,RN1362,claeys:2017b} The norm of \eqref{eq:ABA} is obtained from the determinant
\begin{align}
\braket{\{u\} | \{u\}} &= \det G
\end{align}
with the elements of the so-called Gaudin matrix
\begin{align} \label{eq:GaudinMatrix}
G_{ab} &=
\begin{cases}
 \sum^{K}_{i} \frac{1}{(u_{a}-\varepsilon_{i})^{2}} -2\sum^{M}_{c\neq a}\frac{1}{(u_{a}-u_{c})^{2}} & a=b\\ \frac{2}{(u_{a}-u_{b})^{2}} & a\neq b
\end{cases}.
\end{align}

The normalized 1-RDM may be written
\begin{align} \label{eq:1RDM}
\gamma_{i} = \frac{1}{2} \left(1 - \frac{2^{M}}{\braket{\{u\} | \{u\}}} \frac{\det(Q_i)}{\prod^{M}_{a=1}\prod^M_{b\neq a}(u_a - u_b )}\right)
\end{align}
where the matrix $Q_i$ is defined:
\begin{align}
(Q_i)_{ba} = 
\begin{cases}
\prod^M_{c\neq a} (u_c - u_a) \left(\frac{1}{2} \sum^{K}_{k} \frac{1}{(\varepsilon_{k}-u_{a})^2} -\sum^M_{d\neq a} \frac{1}{(u_d - u_a)^2} - \frac{1}{(\varepsilon_i -u_a)^2}  \right) & a=b \\
\prod^M_{c\neq a} (u_c - u_a) \left(\frac{1}{(u_b - u_a)^2} -\frac{1}{(\varepsilon_i - u_b)^2} \right) & a\neq b
\end{cases}
\end{align}
To arrive at these expressions, the interested reader is referred to ref\cite{RN1586}. 


Unnormalized expressions for both $P_{ij}$ and $D_{ij}$ can be written as sums of determinants related to the Gaudin matrix:

\begin{align}
P_{ij} = \sum^{M}_{a} \frac{u_a - \varepsilon_i}{u_a - \varepsilon_j} \det A^{(i,j)}_{a}
\end{align}
\begin{align}
D_{ij} = -\frac{1}{2}\sum^{M}_{a} \left( \det A^{(i,j)}_{a} + \det A^{(j,i)}_{a} \right) + \frac{1}{2} \left( \gamma_{i} + \gamma_{j} \right)
\end{align}
The matrices $A_a$ appear bizarre at first sight as they are the result of column operations which have condensed a double sum of determinants into a single sum:
\begin{align}
A^{(i,j)}_{a} &= 
\begin{cases}
\vec{G}_{c}-\frac{(\varepsilon_i -u_c)(u_a - u_{c+1})}{(\varepsilon_i - u_{c+1})(u_a - u_c)}\vec{G}_{c+1} & c < a-1 \\
\vec{G}_{c} +\frac{2(\varepsilon_{j}-u_{a})(\varepsilon_{i}-u_{a-1})}{u_{a-1} - u_{a}} \vec{B} & c = a-1 \\
\vec{C} & c = a \\
\vec{G}_{c} & c > a
\end{cases}
\end{align}
where $\vec{G}_{c}$ denotes the \textit{c}th column of the Gaudin matrix Eq. \eqref{eq:GaudinMatrix}, $\vec{B}$ is the column vector:
\begin{align}
\vec{B}_{k} = \frac{(2 u_{k}-\varepsilon_{i}-\varepsilon_{j})}{(u_{k}-\varepsilon_{i})^2(u_{k}-\varepsilon_{j})^2},
\end{align}
and $\vec{C}$ is the column vector:
\begin{align}
\vec{C}_{k} = \frac{1}{(u_{k}-\varepsilon_{i})^2}.
\end{align} 

With explicit expressions for the correlation functions, we can evaluate the energy functional \eqref{eq:su2EnergyExpression} with a cost of $\mathcal{O}(N^6)$: each element of the 2-RDM requires a single summation over determinants, and there are $N^2$ elements to compute. Through optimal book-keeping and storage of computed determinants, it would be possible to improve the scaling, though for our purposes we consider this a dead end. More optimal expressions for the correlation functions exist that we will report in a following publication. Our initial guess for the variational parameters $\{ \varepsilon\}$ was based on the diagonal elements of the 1-electron integrals, perturbed with some random noise. For $g$, we started with a small negative value. While the reduced BCS Hamiltonian has $N+1$ parameters, 2 degrees of freedom are lost to choose the scale and reference point for the energy. Thus we could optimize over $N-1$ parameters, but we found that allowing all $N+1$ parameters to vary led to more robust convergence.
The non-linear relationship between the reduced BCS Hamiltonian parameters $\{ \varepsilon\},g$ and the pair-energies $\{u\}$ suggests that numerical gradients of the energy functional are not an effective tool for minimization. Indeed, we have confirmed this with our preliminary numerical tests. We instead chose to use the Nelder-Mead simplex algorithm\cite{neldermead} which worked effectively. There is always the danger that Nelder-Mead will find the wrong optimum, though we have eliminated this issue by preconditioning with the covariance matrix adaptation evolution strategy \cite{cma}.

\section{Numerical Results} \label{sec:numbers}
Calculations were performed for a series of four-, six-, eight-, and ten-electron atomic systems as well as for dissociation of hydrogren chains and molecular nitrogen. Results are compared with doubly-occupied configuration interaction (DOCI)\cite{RN1702} and full configuration interaction (CI). Full CI calculations were performed with psi4\cite{psi4,detci} and verified with an in-house code, DOCI calculations were performed with an in-house code, and RHF calculations were performed with Gaussian 16\cite{g16}. We have noted previously that seniority-zero wavefunction ans\"{a}tze favour localized, valence-bond-like orbitals, rather than the delocalized orbitals obtained from RHF. Thus, dissociation curves were computed both in the basis of RHF orbitals and the basis of orbital optimized DOCI (OO-DOCI) orbitals. Orbital optimization in the OO-DOCI calculations was performed as MC-SCF calculations in the complete doubly-occupied Slater determinant basis with a Newton-Raphson scheme for the optimization for the determinant and orbital coefficients as implemented in GAMESS(US)\cite{GAMESS}. To explore the orbital optimization space more profoundly, several starting bases were constructed including the RHF orbitals, FCI natural orbitals and random orbitals obtained from rotating the other bases. 

The variational RG results should always be compared with DOCI: when computed in the same set of orbitals, RG is a strictly variational approximation to DOCI. For atoms, calculations were performed with STO-6G and with aug-cc-pVDZ, while dissociation curves were computed with STO-6G. As our results are proof of principle, our algorithm is not optimal, which unfortunately limits the size of system we can treat. However, results with STO-6G will isolate effects of strong-correlation. Effects of weak correlation are minimal in STO-6G as there are limited weak excitations possible. Thus, dissociation curves computed with STO-6G are meaningful and relevant.

\subsection{Atoms}
Raw energetic results for atomic systems are reported in Table \ref{sto_absdata} (STO-6G) and Table \ref{aug_absdata} (aug-cc-pVDZ). Each atomic system considered was necessarily treated as a closed-shell singlet. All calculations are performed with the RHF orbitals. Again, for a given basis, in this case RHF orbitals, the best possible result in the space of seniority-zero wavefunctions is DOCI. Thus, we summarize the deviations from DOCI in table \ref{doci_dev}.

\begin{table}[h]
\centering
\begin{tabular}{|l|r|r|r|r|r|r|r|}
\thead{a)} & \thead{Be} & \thead{B$^+$} & \thead{C$^{2+}$} & \thead{N$^{3+}$} & \thead{O$^{4+}$} & \thead{F$^{5+}$} & \thead{Ne$^{6+}$}  \\
\hline  
RHF     & -14.50336 & -24.19056 & -36-34155 & -50.87786 & -67.89148 & -87.35546 & -109.32595   \\
RG      & -14.55578 & -24.25254 & -36.40430 & -50.94130 & -67.95846 & -87.42542 & -109.39974   \\
DOCI    & -14.55578 & -24.25254 & -36.40430 & -50.94130 & -67.95847 & -87.42542 & -109.39974   \\
FCI     & -14.55609 & -24.25289 & -36.40457 & -50.94153 & -67.95870 & -87.42566 & -109.40001   \\
\hline
\end{tabular}
\begin{tabular}{|l|r|r|r|r|r|r|r|}
\thead{b)} & \thead{Be$^{2-}$} & \thead{B$^{-}$} & \thead{C} & \thead{N$^{+}$} & \thead{O$^{2+}$} & \thead{F$^{3+}$} & \thead{Ne$^{4+}$}  \\
\hline  
RHF     & -13.61385 & -24.01092 & -37.46352 & -53.64180 & -72.65698 & -94.54252 & -119.37771 \\
RG      & -13.65525 & -24.06267 & -37.52018 & -53.70354 & -72.72618 & -94.61900 & -119.46229 \\
DOCI    & -13.65525 & -24.06267 & -37.52018 & -53.70356 & -72.72618 & -94.61900 & -119.46238 \\
FCI     & -13.70391 & -24.12611 & -37.59286 & -53.78592 & -72.82119 & -94.72667 & -119.58397 \\
\hline
\end{tabular}
\begin{tabular}{|l|r|r|r|r|r|r|r|}
\thead{c)} & \thead{Be$^{4-}$} & \thead{B$^{3-}$} & \thead{C$^{2-}$} & \thead{N$^{-}$} & \thead{O} & \thead{F$^{+}$} & \thead{Ne$^{2+}$}  \\
\hline  
RHF     & -11.16645 & -21.79925 & -36.25543 & -53.76411 & -74.37443 & -98.27513 & -125.52797 \\
RG      & -11.19071 & -21.83084 & -36.29171 & -53.80525 & -74.42158 & -98.32892 & -125.58872 \\
DOCI    & -11.19071 & -21.83089 & -36.29171 & -53.80525 & -74.42189 & -98.32892 & -125.58872 \\
FCI     & -11.23923 & -21.89417 & -36.36427 & -53.88751 & -74.51682 & -98.43650 & -125.71022 \\
\hline
\end{tabular}
\caption{\label{sto_absdata} Absolute energies (a.u.) computed with the STO-6G basis set for a) four-electron systems, b) six-electron systems and c) eight-electron systems.}
\end{table}

\begin{table}[h]
\centering
\begin{tabular}{|l|r|r|r|r|r|r|r|}
\thead{a)}& \thead{Be} & \thead{B$^+$} & \thead{C$^{2+}$} & \thead{N$^{3+}$} & \thead{O$^{4+}$} & \thead{F$^{5+}$} & \thead{Ne$^{6+}$}  \\
\hline  
RHF     & -14.57238 & -24.23501 & -36.40165 & -51.06854 & -68.23528 & -87.89994 & -110.06183  \\
RG      & -14.59411 & -24.27572 & -36.46351 & -51.14659 & -68.32741 & -88.00447 & -110.17753  \\
DOCI    & -14.59430 & -24.27614 & -36.46414 & -51.14733 & -68.32777 & -88.00475 & -110.17811  \\
FCI     & -14.61747 & -24.29450 & -36.47469 & -51.15446 & -68.33310 & -88.00914 & -110.18197  \\
\hline
\end{tabular}
\begin{tabular}{|l|r|r|r|r|r|r|r|}
\thead{b)} & \thead{Be$^{2-}$} & \thead{B$^{-}$} & \thead{C} & \thead{N$^{+}$} & \thead{O$^{2+}$} & \thead{F$^{3+}$} & \thead{Ne$^{4+}$}  \\
\hline  
RHF     & -14.42518 & -24.47453 & -37.59848 & -53.75628 & -72.92427 & -95.09707 & -120.26904 \\
RG      & -14.44946 & -24.50437 & -37.62045 & -53.80071 & -72.98480 & -95.17516 & -120.36321 \\
DOCI    & -14.45682 & -24.51240 & -37.62790 & -53.80695 & -72.99078 & -95.17980 & -120.36782 \\
FCI     & -14.50771 & -24.57900 & -37.71348 & -53.88304 & -73.06314 & -95.24915 & -120.43495 \\
\hline
\end{tabular}
\begin{tabular}{|l|r|r|r|r|r|r|r|}
\thead{c)} & \thead{Be$^{4-}$} & \thead{B$^{3-}$} & \thead{C$^{2-}$} & \thead{N$^{-}$} & \thead{O} & \thead{F$^{+}$} & \thead{Ne$^{2+}$}  \\
\hline  
RHF     & -13.94252 & -24.00987 & -37.35787 & -54.22984 & -74.67005 & -98.64114 & -126.12742 \\
RG      & -13.96797 & -24.04881 & -37.40351 & -54.25914 & -74.70246 & -98.68616 & -126.18290 \\
DOCI    & -13.97955 & -24.06759 & -37.42464 & -54.27194 & -74.71325 & -98.69661 & -126.19256 \\
FCI     & -14.10357 & -24.15193 & -37.54951 & -54.41593 & -74.84971 & -98.82188 & -126.31146 \\
\hline
\end{tabular}
\begin{tabular}{|l|r|r|r|r|r|r|r|}
\thead{d)}& \thead{Be$^{6-}$} & \thead{B$^{5-}$} & \thead{C$^{4-}$} & \thead{N$^{3-}$} & \thead{O$^{2-}$} & \thead{F$^{-}$} & \thead{Ne}  \\
\hline  
RHF     & -13.14242 & -23.11023 & -36.43155 & -53.48638 & -74.43570 & -99.42828 & -128.49635 \\
RG      & -13.16941 & -23.14395 & -36.48787 & -53.54620 & -74.47711 & -99.46036 & -128.52650 \\
DOCI    & -13.17847 & -23.16554 & -36.51415 & -53.59230 & -74.50614 & -99.48046 & -128.54457 \\
FCI     & -13.41990 & -23.25556 & -36.67971 & -53.78809 & -74.72569 & -99.67132 & -128.71147 \\
\hline
\end{tabular}
\caption{\label{aug_absdata} Absolute energies (a.u.) computed with the aug-cc-pVDZ basis set for a) four-electron systems, b) six-electron systems, c) eight-electron systems and d) ten-electron systems.}
\end{table}

\begin{table}[h]
\centering
\begin{tabular}{|l|r|r|r|r|r|r|r|}
\thead{a)}& \thead{Be} & \thead{B} & \thead{C} & \thead{N} & \thead{O} & \thead{F} & \thead{Ne}  \\
\hline  
4e      & 1.94E-6 & 1.43E-6 & 5.47E-7 & 2.53E-7 & 3.30E-7 & 8.86E-8 & 3.15E-7 \\
6e      & 2.20E-7 & 5.93E-7 & 2.98E-8 & 2.34E-5 & 1.07E-7 & 8.87E-7 & 8.58E-5 \\
8e      & 8.33E-8 & 4.89E-5 & 1.10E-8 & 2.58E-8 & 3.17E-4 & 4.68E-8 & 6.78E-7 \\
\hline
\end{tabular}
\begin{tabular}{|l|r|r|r|r|r|r|r|}
\thead{b)}& \thead{Be} & \thead{B} & \thead{C} & \thead{N} & \thead{O} & \thead{F} & \thead{Ne}  \\
\hline  
4e      & 1.88E-4 & 4.23E-4 & 6.30E-4 & 7.46E-4 & 3.59E-4 & 2.74E-4 & 5.74E-4 \\
6e      & 7.36E-3 & 8.03E-3 & 7.44E-3 & 6.24E-3 & 5.98E-3 & 4.64E-3 & 4.61E-3 \\
8e      & 1.16E-2 & 1.88E-2 & 2.11E-2 & 1.28E-2 & 1.08E-2 & 1.04E-2 & 9.67E-3 \\
10e     & 9.06E-3 & 2.16E-2 & 2.08E-2 & 2.88E-2 & 2.90E-2 & 2.01E-2 & 1.81E-2 \\
\hline
\end{tabular} 
\caption{\label{doci_dev} Deviations of variational RG with DOCI computed in the a) STO-6G and b) aug-cc-pVDZ basis sets.}
\end{table}

For the four-electron series there is one pair of electrons deeply entrenched in the 1s core, while the second pair resides principally in the 2s spatial orbital. The 2s-2p gap shrinks as the central charge becomes more positive, and thus the electronic configurations with the second pair occupying the 2p spatial orbitals become important, which makes these systems strongly correlated. All the important Slater determinants in the physical wavefunction are seniority zero, so DOCI is near-exact, as are our variational RG results. 

In the six-electron series there are two pairs of electrons in the valence orbitals, and the systems are once again strongly-correlated. However, DOCI is not as good a treatment, as there is weak-correlation from open-shell singlet states missing from DOCI. The same is generally true for the results of the eight-electron series. 

For the ten-electron series the dominant effect is weak electron correlation. The 2s-2p gap again gets smaller as the central charge increases, but the 2p-3s gap remains large, and thus there is a single Slater determinant which dominates the physical wavefunction. As a result, DOCI is not quantitatively accurate and neither is our variational calculation. We do not report results for STO-6G as HF is full CI for this case.

In each case we were able to reproduce from half to two-thirds of the correlation energy obtainable by DOCI, which is the best-case scenario for this wavefunction ansatz. To recover the complete DOCI correlation energy, one way to proceed is to write an expansion in terms of eigenvectors of the reduced BCS Hamiltonian. As we already recover the majority of the correlation energy, we are optimistic that such an expansion is short, and dominated by \emph{one} RG state. We are now pursuing this line of reasoning and will report in the future.

\subsection{Dissociation curves: RHF orbitals}

\begin{figure}[h]
	\centering
	\begin{subfigure}[b]{0.475\textwidth}
		\centering
		\includegraphics[width=\textwidth]{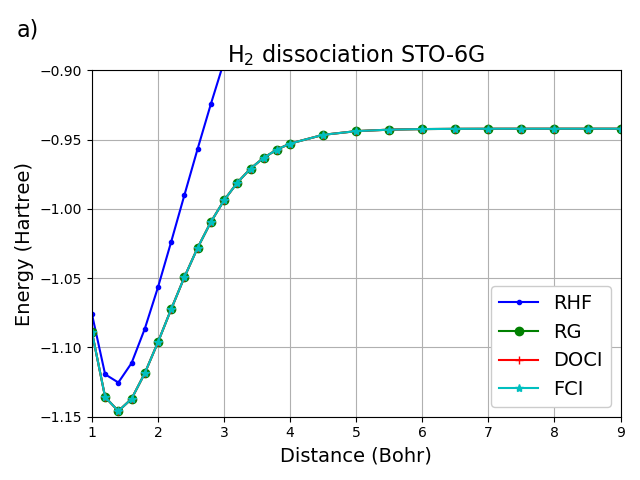}
	\end{subfigure}
	\hfill
	\begin{subfigure}[b]{0.475\textwidth}
		\centering
		\includegraphics[width=\textwidth]{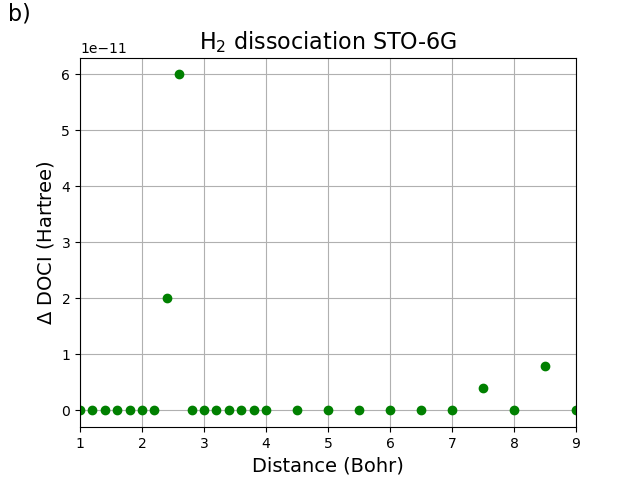}
	\end{subfigure}
	\caption{a) Bond dissociation curves for H$_2$. RG and DOCI were computed in the basis of RHF orbitals. RG, DOCI and FCI coincide and hence are not distinguishable. b) Energy difference between RG and DOCI for H$_2$. All results were computed with the STO-6G basis set.}
	\label{H2_curves}
\end{figure}

The prototypical strongly-correlated systems are bond-dissociation curves. As it is a two-electron problem, we expect to be able to dissociate the hydrogen molecule perfectly. Indeed this is the case, as can be seen in Figure \ref{H2_curves}, where the RG, DOCI, and FCI curves overlap. The error with respect to DOCI is very small.

\begin{figure}[h]
	\begin{subfigure}[b]{0.475\textwidth}
		\centering
		\includegraphics[width=\textwidth]{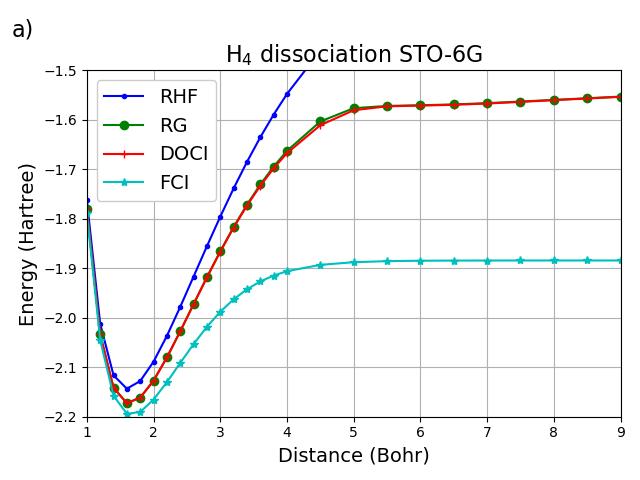}
	\end{subfigure}
	\hfill
	\begin{subfigure}[b]{0.475\textwidth}
		\centering
		\includegraphics[width=\textwidth]{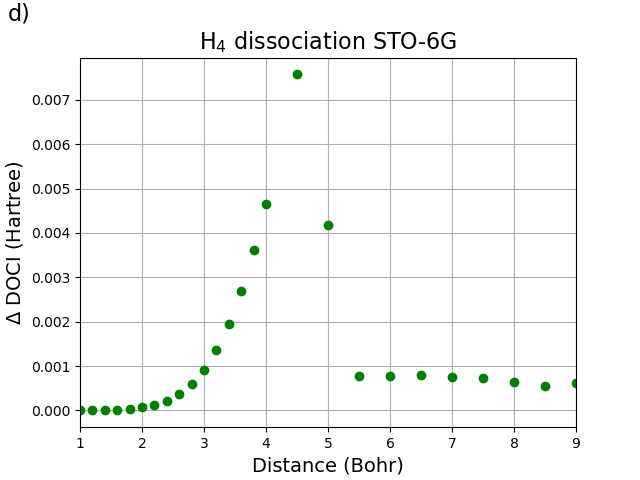}
	\end{subfigure}
	\begin{subfigure}[b]{0.475\textwidth}
		\centering
		\includegraphics[width=\textwidth]{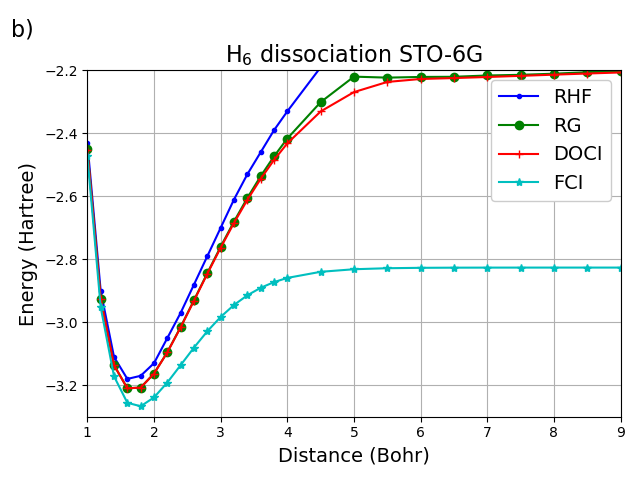}
	\end{subfigure}
	\hfill
	\begin{subfigure}[b]{0.475\textwidth}
		\centering
		\includegraphics[width=\textwidth]{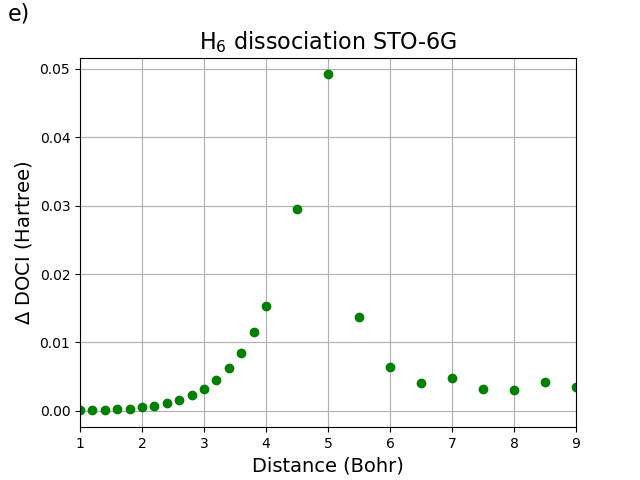}
	\end{subfigure}
	\begin{subfigure}[b]{0.475\textwidth}
		\centering
		\includegraphics[width=\textwidth]{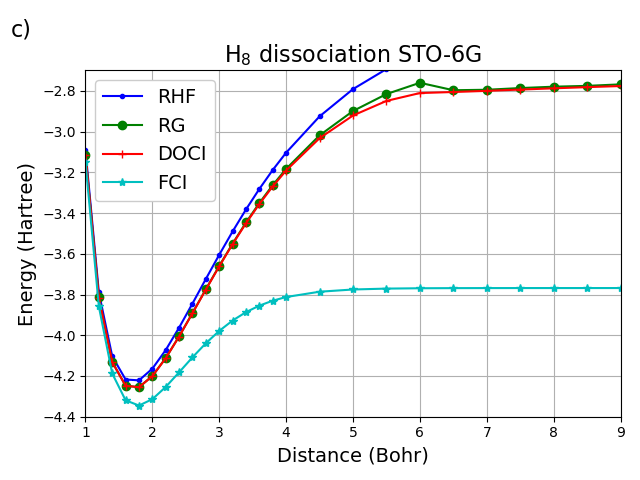}
	\end{subfigure}
	\hfill
	\begin{subfigure}[b]{0.475\textwidth}
		\centering
		\includegraphics[width=\textwidth]{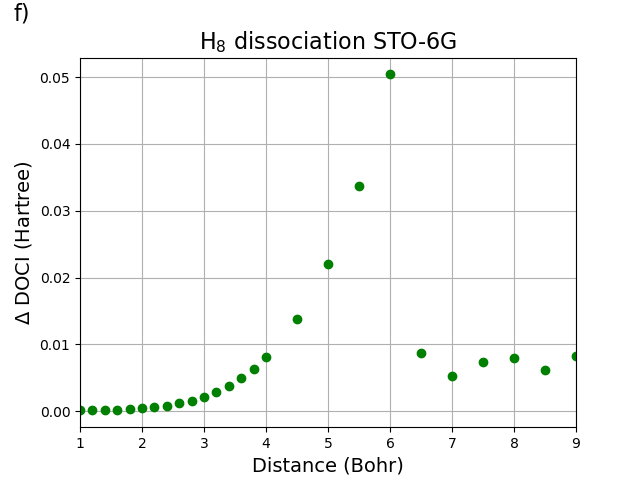}
	\end{subfigure}
	\caption{a-c) Bond dissociation curves for H$_4$, H$_6$ and H$_8$. d-f) Energy difference between RG and DOCI for H$_4$, H$_6$ and H$_8$. All results were computed with the STO-6G basis set. RG and DOCI were computed in the basis of RHF orbitals.}
		\label{H_chain_curves}
\end{figure}

Moving to the simultaneous dissociation of linear H$_4$ into four hydrogen atoms, as shown in figure \ref{H_chain_curves}, the results are no longer exact. As all calculations are in the RHF basis, DOCI and FCI differ appreciably. The error for RG with respect to DOCI is no longer zero, but grows continuously to a maximum before dropping off substantially. At the critical point, where the deviation is maximal, more than one RG state is required to match with DOCI. The same trends are observed for H$_6$ and H$_8$.

\subsection{Dissociation curves: OO-DOCI orbitals}
\begin{figure}[h]
	\begin{subfigure}[b]{0.475\textwidth}
		\centering
		\includegraphics[width=\textwidth]{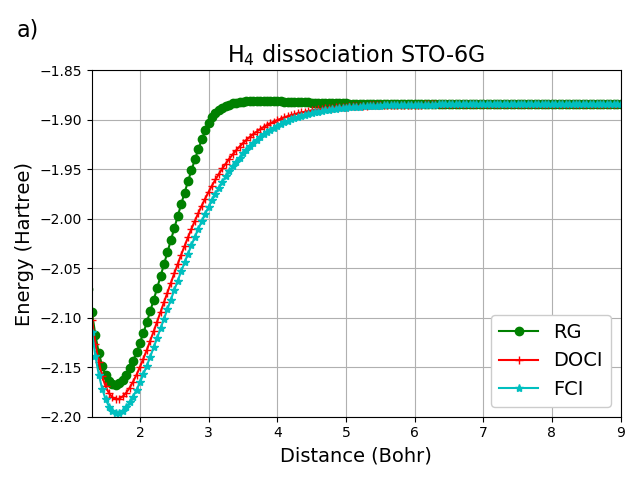}
	\end{subfigure}
	\hfill
	\begin{subfigure}[b]{0.475\textwidth}
		\centering
		\includegraphics[width=\textwidth]{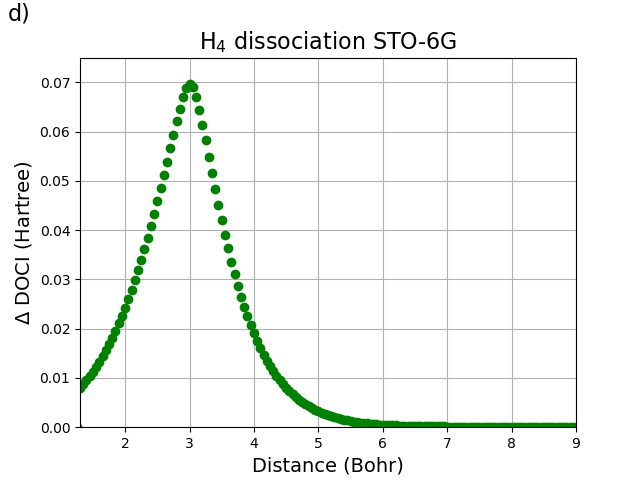}
	\end{subfigure}

	\begin{subfigure}[b]{0.475\textwidth}
		\centering
		\includegraphics[width=\textwidth]{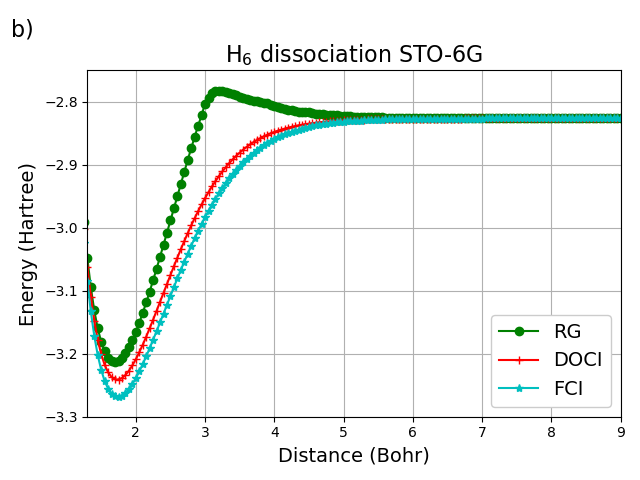}
	\end{subfigure}
	\hfill
	\begin{subfigure}[b]{0.475\textwidth}
		\centering
		\includegraphics[width=\textwidth]{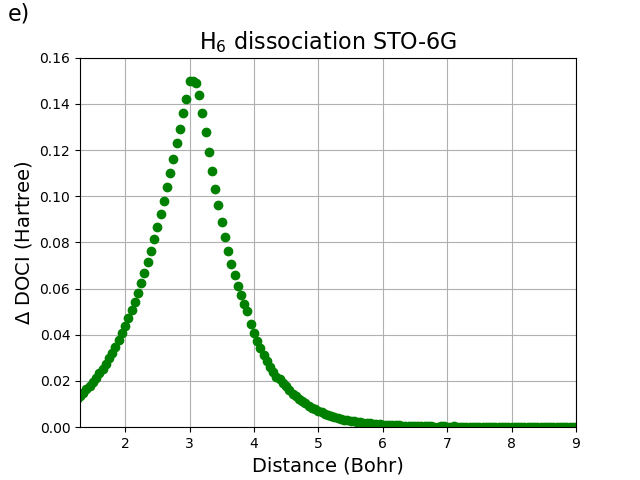}
	\end{subfigure}
	
	\begin{subfigure}[b]{0.475\textwidth}
		\centering
		\includegraphics[width=\textwidth]{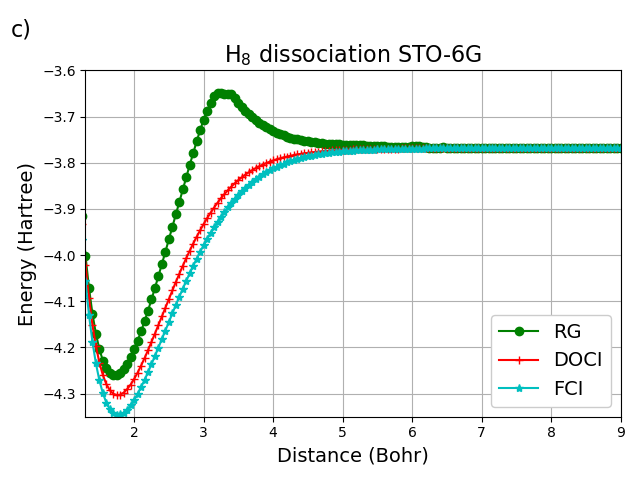}
	\end{subfigure}
	\hfill
	\begin{subfigure}[b]{0.475\textwidth}
		\centering
		\includegraphics[width=\textwidth]{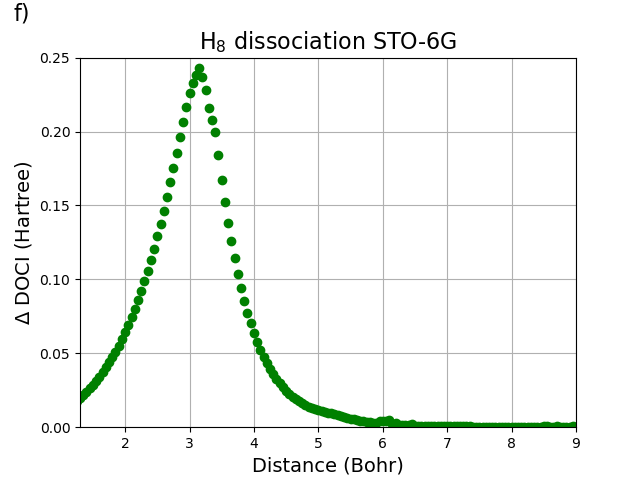}
	\end{subfigure}
	\caption{a-c) Bond dissociation curves for H$_4$, H$_6$ and H$_8$. d-f) Energy difference between RG and DOCI for H$_4$, H$_6$ and H$_8$. All results were computed with the STO-6G basis set. RG and DOCI were computed in the basis of OO-DOCI orbitals.}
		\label{H_chain_DOCI_curves}
\end{figure}

Hydrogen dissociation chains were also computed in the basis of OO-DOCI orbitals, in which the DOCI curve is much closer to the full CI result. Curves are plotted for H$_4$, H$_6$ and H$_8$ in figure \ref{H_chain_DOCI_curves}. The results for each of the hydrogen chains are the same. The error with respect to DOCI in the RG curve grows continuously before decreasing to less than 1mH at dissociation. That the error tends to zero is a strong indication that the method is size-consistent.

Dissociation curves were also calculated for the nitrogen molecule, and are plotted in figure \ref{N2_DOCI_curves}. Similar to the case for hydrogen chains, RG differs from DOCI near the minimum, but approaches the DOCI curve much more quickly. There is a curve crossing near 5.2 Bohr which indicates that there is more than one RG state required near that point. At dissociation the RG and DOCI curves agree to a tenth of a milliHartree.
\textbf{\begin{figure}[h]
	\begin{subfigure}[b]{0.475\textwidth}
		\centering
		\includegraphics[width=\textwidth]{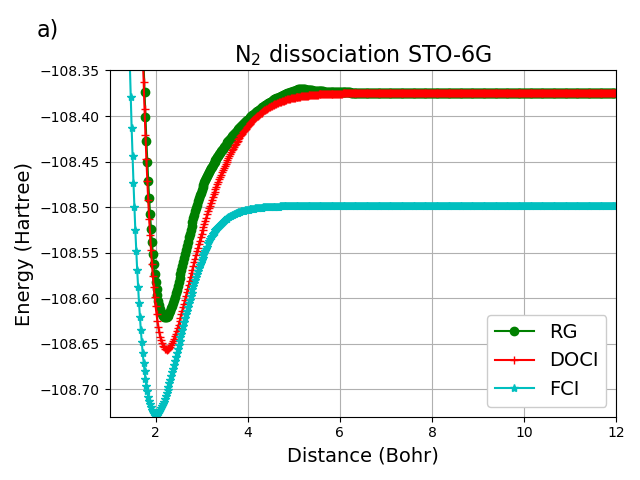}
	\end{subfigure}
	\hfill
	\begin{subfigure}[b]{0.475\textwidth}
		\centering
		\includegraphics[width=\textwidth]{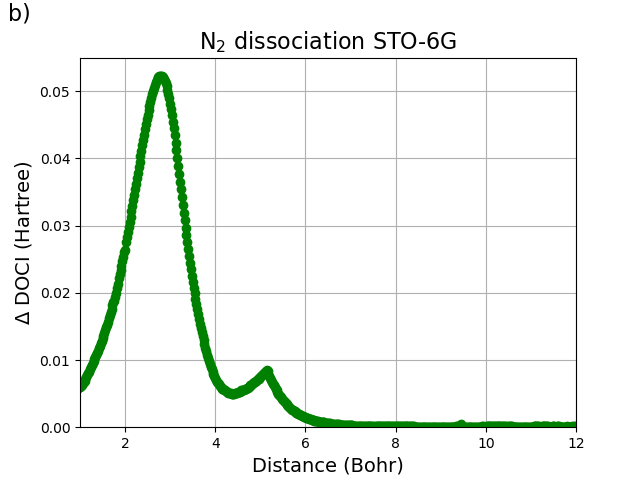}
	\end{subfigure}
	\caption{a) Bond dissociation curves for N$_2$. b) Energy difference between RG and DOCI for N$_2$. All results were computed with the STO-6G basis set. RG and DOCI were computed in the basis of OO-DOCI orbitals.}
		\label{N2_DOCI_curves}
\end{figure}}

\section{Conclusions} \label{sec:conclusions}

We have performed variational calculations for chemical systems employing the ground state eigenvector of the exactly solvable reduced BCS Hamiltonian. The key idea is that this treatment is a mean-field of pairs of electrons, rather than a mean-field of individual electrons, as in conventional orbital-based approaches. Analogous to the way Hartree-Fock is the dominant contribution to the wavefunction of a system with weakly-correlated electrons, the present method is the dominant contribution to a wavefunction of a system with weakly-correlated pairs of electrons. 

Our results serve as a starting point to develop a many-body theory for pairs of electrons. We are satisfied that they qualitatively reproduce DOCI. They also highlight issues to be addressed in upcoming contributions. It is obvious that RHF orbitals are not optimal for seniority-zero wavefunctions, as we have studied previously. Weak-correlation of pairs, or inter-pair correlation, is missing, and perturbation theories will need to be developed. Finally, while our method scales polynomially, it should be a smaller polynomial to be taken seriously. All of these problems are solvable, and we are currently addressing them. 

\section{Data Availability}
The data that support the findings of this study are available from the corresponding author upon reasonable request.

\section{Acknowledgements}
P.A.J and P.W.A were supported by NSERC and Compute Canada. P.W.A. also thanks Canarie. P.W.A and S. D. B. thank the Canada Research Chairs. We thank Toon Verstraelen for his implementation of the covariance matrix adaptation evolution strategy, and Pieter Claeys for his Richardson's equations solver.

\appendix
 
\bibliography{on_shell_clean}

\bibliographystyle{unsrt}

\end{document}